\documentclass[pss]{wiley2sp} 
\usepackage{amsmath}

\def\be{\begin{equation}}
\def\ee{\end{equation}}

\begin{document}

\title{Theoretical study of hydrogen microstructure in models of hydrogenated amorphous silicon}

\titlerunning{ Hydrogen microstructure of a-Si:H}

\author{%
Rajendra Timilsina \textsuperscript{\textsf{\bfseries}}, Parthapratim Biswas \textsuperscript{\Ast}}

\authorrunning{Timilsina et al.}

\mail{e-mail
  \textsf{Partha.Biswas@usm.edu}, Phone:
  +1-601-2665156, Fax: +1-601-2665149}

\institute{Department of Physics and Astronomy, The University of Southern Mississippi, Hattiesburg, MS 39406, USA \\
}

\received{XXXX, revised XXXX, accepted XXXX} 
\published{XXXX} 


\abstract{%
%
%
%
\abstcol{%
We study the distribution of hydrogen and various hydride configurations in
realistic models of a-Si:H for two different concentration generated via experimentally
constrained molecular relaxation approach (ECMR)~\cite{ecmr}. The microstructure corresponding
to low ( $<$ 10\%) and high ($>$ 20\%) concentration of H atoms are addressed and
are compared to the experimental results with particular emphasis on the size of H 
clusters  and local environment of H atoms.}{The linewidths of the nuclear magnetic
resonance (NMR) spectrum for the model configurations are calculated in order
to compare to the experimental NMR data. Our study shows the presence
of isolated hydrogen atoms, small and relatively large clusters with average proton-proton
neighbor distance in the clusters around 1.6 - 2.4 {\AA} that have been observed in
multiple quantum NMR experiments.}}


\maketitle   

\section{ Introduction}

Hydrogenated amorphous silicon (a-Si:H) is a very important material with a wide
variety of technological applications ranging from photovoltaics to memory switching
devices, thin film transistors, solar cells, optical scanners and numerous other
electronic instruments~\cite{Street}.
The material has been studied extensively starting from its structural, electronic to
vibrational and optical properties that produced a large volume of experimental data.
In spite of this, our understanding of a-Si:H from theoretical
point of view is very limited. Some of the very fundamental processes involving
the material are poorly understood. For example, the creation of
metastable defects in the material upon prolonged exposure to light by breaking
of a Si--Si bond next to Si--H bond, the so-called Staebler-Wronski effect (SW)~\cite{Staebler}, 
is perhaps the most intriguing problem that is yet to be understood. Since many of
the proposed microscopic mechanisms that attempt to explain the SW effect are based
on bond-breaking models~\cite{sw-model1} \cite{sw-model2}, it is crucial to understand the distribution
of H atoms and their dynamics in the amorphous network. Although there exists a number
of experimental studies on hydrogen distribution in the material via nuclear magnetic
resonance (NMR)~\cite{Reimer}\cite{Baum}\cite{Rutland} and infrared (IR) spectroscopy\cite{ir}, there are very 
few theoretical studies that address the problem explicitly~\cite{Biswas}\cite{Drabold}.

In this paper we propose to calculate the hydrogen microstructure of realistic models of a-Si:H
at low and high concentration.  Since one-dimensional NMR data cannot be uniquely
mapped onto real-space distribution without the uncertainty of various fitting
procedures, it is much more useful and convenient to employ first-principles models
of a-Si:H to generate a distribution of hydrogen atoms.  The NMR linewidths of
the distribution can be obtained from a knowledge of the position of the spins in the network using
a suitable approximation on the shape of the linespectra. This approach provides a direct route 
to study the number and the size of hydrogen clusters, their local environment, and the effect 
of the cluster size on the NMR linewidths.  The dependence of H microstructure on the 
concentration of H atoms is also addressed and a comparison is made to experimental data. 
The plan of the paper is as follows. In section 2, we summarize
the various experimental results from NMR experiments. Section 3 then describes briefly the a-Si:H
models, and the resulting distribution of hydrogens at two different concentration with emphasize
on sparse and clustered environment. In section 4, we calculate the NMR linewidths
for the models and compared to experimental data, which is then followed by conclusion of
our work.

\section{Summary of experimental results}
The two principal methods that are used to obtain experimental data on hydrogen
distribution in a-Si:H are nuclear magnetic resonance (NMR)\cite{Reimer}\cite{Baum}\cite{Rutland} 
and infrared spectroscopy (IR)\cite{ir}. The former provides useful information about the local environment of H atoms, whereas
the latter plays a dominant role in determining the nature of various silicon-hydrogen bonding
configurations.  Experimental data suggest that H atoms can reside both in isolated and clustered phase
and that the distribution can be highly inhomogeneous \cite{Rutland} \cite{Schropp}, which
depends on a number of factors such as preparation conditions, method of deposition,
substrate temperature etc.  A typical NMR spectrum of a device-quality sample shows the
presence of both narrow and broad linewidths, which can be approximated as a convolution of
a Lorentzian (narrow) and Gaussian (broad) distribution. The dipolar interaction between the protons 
suggests that the narrow width is associated with a dilute or sparse distribution of spins, whereas 
a dense or clustered environment is needed to produce a broad linewidth in the resonance spectrum.
Baum et al.~\cite{Baum} demonstrated via multiple-quantum NMR experiments that for device-quality 
samples the clustering of H atoms increases with the increase of H concentration.  The typical size 
of the clusters found in their work was of the order of 4-6 atoms, and with increasing concentration 
the clusters became physically closer to one another. NMR and IR  measurements by Gleason et 
al.~\cite{Gleason} of a-Si:H prepared via plasma-enhanced chemical vapor deposition (PECVD) also confirmed 
this observation. These authors reported the presence of isolated, small and medium size clusters in 
the samples that gave rise to both broad and narrow linewidths of about 28-33 kHz and  6-12 kHz respectively 
in the concentration range 10 - 16\%. In other words, the picture that emerges from NMR and IR experiments 
on samples prepared by a variety of methods is that for device-quality samples H prefers to be distributed 
sparsely in the sample with occasional clustering at low temperature.
While the clustering is more likely at high concentration, the
microstructure can be very inhomogeneous, and may consist of few large clusters even at very
low concentration. An example of such inhomogeneous distribution of hydrogen atoms was reported
by Wu et al. in hot-filament-assisted CVD deposited a-Si:H film \cite{Rutland}.
They observed a significantly large width of about 50 kHz in HW sample at very low concentration (2-3\%)
consisting of at least one  H complex that was quite different from device-quality glow-discharged
a-Si:H at 8-10\%. This broad width is an indicative of the presence of large clusters 
in the sample that makes the distribution very inhomogeneous even at very low 
concentration.

\section{ Microstructure from model a-Si:H}
In order to study hydrogen microstructure, we have used two realistic models of a-Si:H obtained via
experimentally constrained molecular relaxation approach (ECMR)\cite{ecmr}. 
The method is based on a hybrid approach that has two important components: 1) experimental data can be
directly utilized in building the model configurations, and 2) the model configurations are then driven 
via first-principles force-fields to obtain the minimum energy configuration. 
The method is unique in the sense that it is able to avoid some of the common pitfalls of
both traditional first-principle molecular dynamics (small size, and large simulation time), and the
uncertainty associated with constructing models solely based on limited experimental data (such as reverse 
Monte Carlo~\cite{rmc}). We generate two models 
for our purpose of microstructure calculation. 
The first model consists of 540 atoms with 7\% H atoms, while the second one has 611 atoms with 
22\% H atoms. The structural, electronic and vibrational properties of the models had been extensively 
studied by one of us and were compared to experimental data obtained from X-ray and Neutron diffraction 
measurements\cite{ecmr-H2}. Since the microstructure can be very sensitive to the details of the 
model construction, it is important to validate the properties of model networks with experimental data 
prior to addressing hydrogen microstructure. 

\begin{figure}[t]
\includegraphics[width=2.8 in, height=2.3 in, angle =0 ]{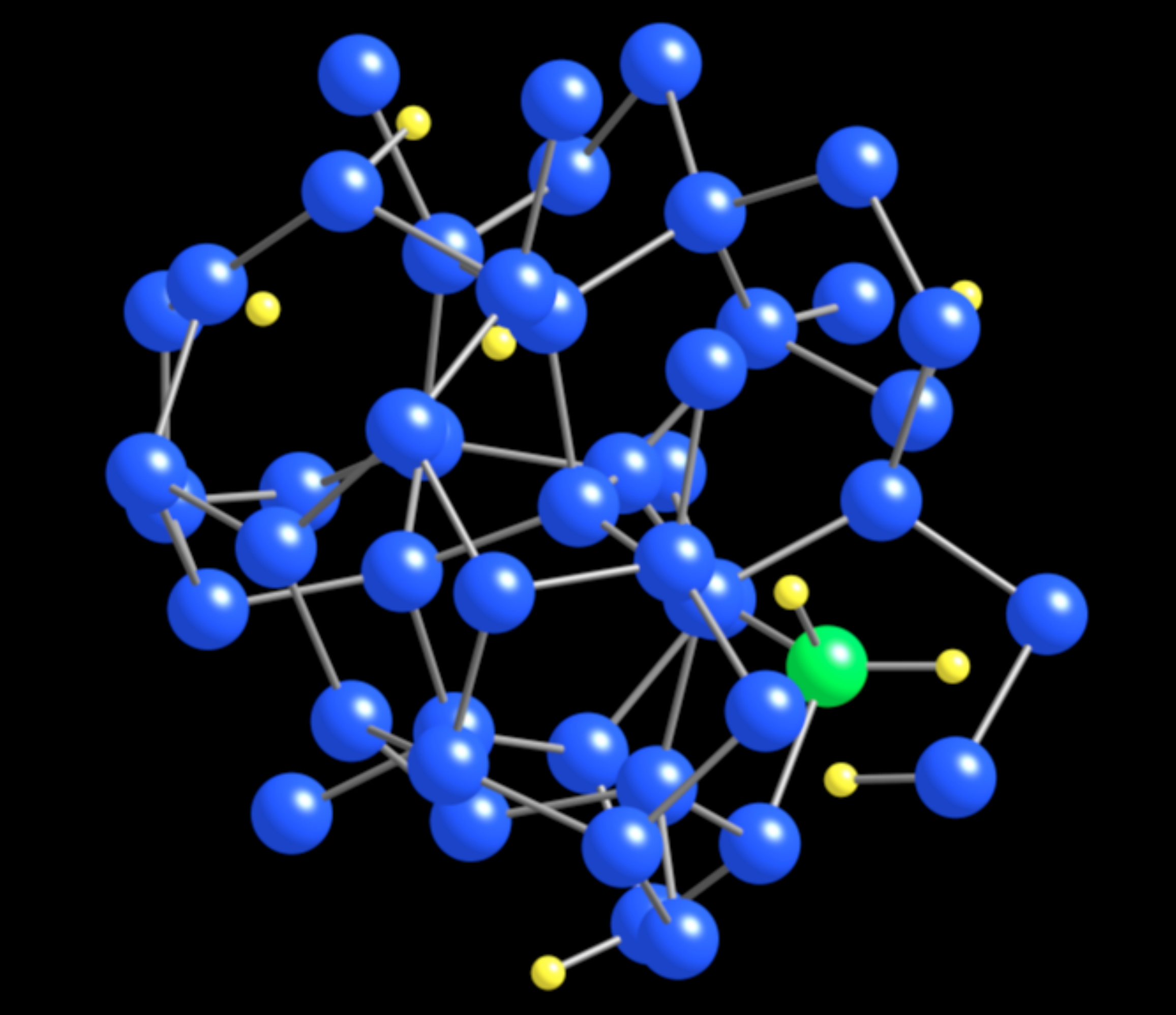}
\caption{
\label{fig1}
(Color online)
Hydrogen distribution and its local environment in a model of a-Si:H with 7\% H atoms. 
The
distribution of H atoms (in yellow/white) is sparse showing isolated, monohydride and dihydride
configurations.  The Si atoms are shown in blue/black with the exception of the one in 
the dihydride configuration (in green/white).
}
\end{figure}

The 540-atom model with 7\% H atoms has very few defects and closely corresponds to a device-quality sample
at low concentration. The model has a clean electronic band gap and the correct vibrational density of states
showing the characteristic acoustic and optical peaks that have been discussed at length in Ref.\cite{ecmr-H2}.
A real-space analysis of this model shows the presence of both isolated and clustered phase of H atoms. In fig.\ref{fig1}
we have shown the distribution of hydrogens in a region of linear dimension 6 {\AA} that consists of 6-7 H atoms.
The presence of isolated H atom, Si-H and SiH$_2$ configurations are also visible in the figure. An SiH$_2$
configuration is explicitly shown with the Si atom in green/white. 
The local environment of this Si atom shows clustering of H atoms within the radius of 3-4 {\AA}.
The model consists of several such regions some of which have more dilute or dispersed atoms, while the others
are relatively less sparse with small clusters of 4-5 atoms. The proton-proton distance in this model is
approximately lying between 1.6 {\AA} to 2.4 {\AA}. Our analysis shows that although there are few clustered 
phase (within 4-6 {\AA}), there are regions where the H atoms are separated by relatively large distances.
The overall microstructure of the model is shown in fig. \ref{fig3} without the embedding Si matrix
(except silicon-hydrogen bonding) in order to highlight the sparse and densely occupied regions.
It is evident that H distribution is quite sparse (except for few small clusters) consisting of few 
isolated H atoms, monohydride SiH and dihydride SiH$_2$ configurations. Gleason et al.~\cite{Gleason}  
studied a-Si:H films obtained by plasma enhanced CVD at various temperature
in the concentration range 9 - 16 \% of hydrogen via multiple quantum NMR (MQ-NMR) and suggested the presence of small clusters
in the samples. Baum et al.~\cite{Baum} also arrived at the similar conclusion studying device-quality samples via
MQ-NMR in their works and reported the presence of cluster of 4 to 7 atoms. These observations are consistent with the results
obtained from our model network. The implication of disperse and clustered environment of H atoms in the network on 
NMR linewidths is discussed in the next section.

\begin{figure}[t]
\includegraphics[width = 2.7 in,height= 2.3in , angle =0 ]{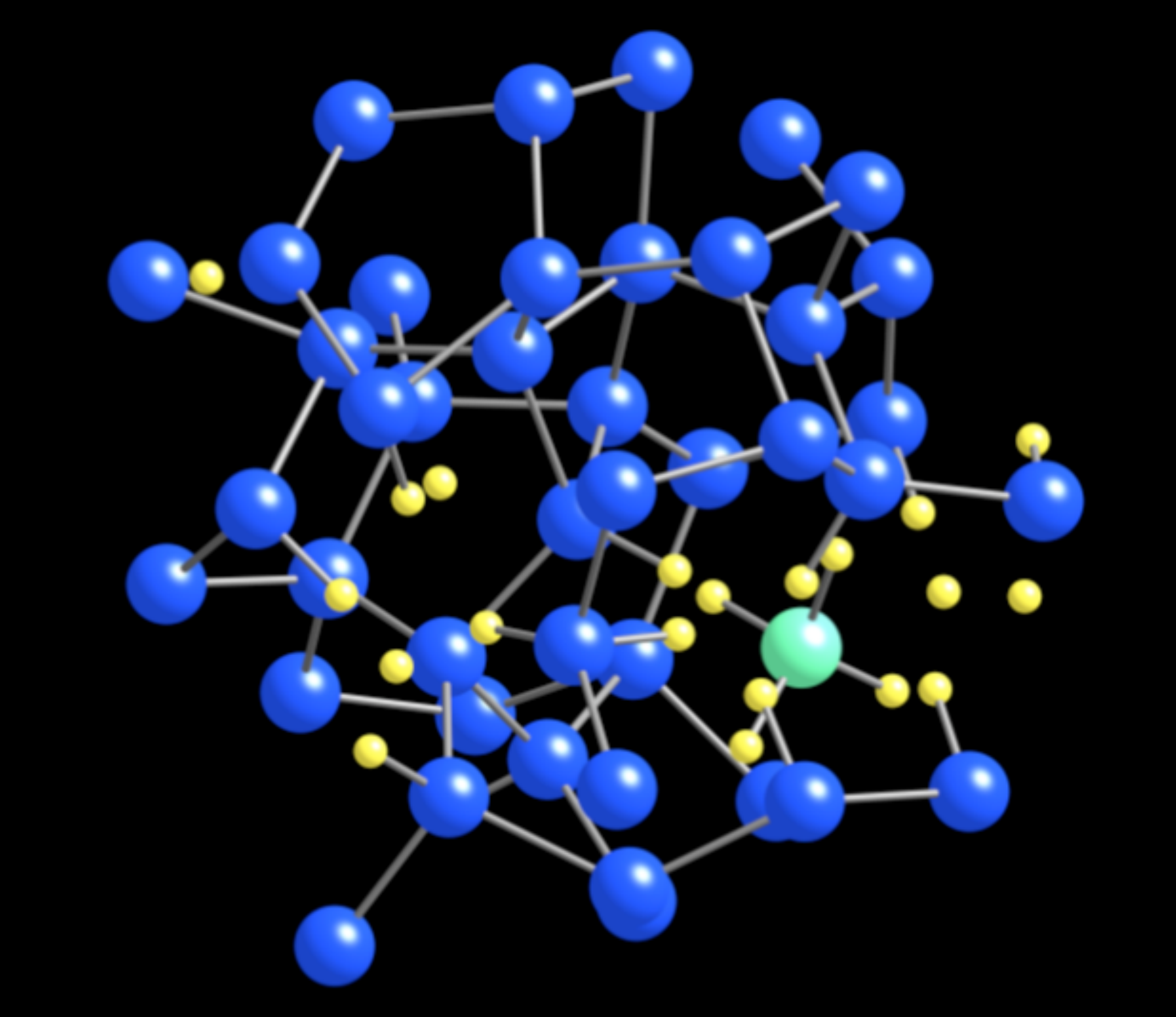}
\caption{
\label{fig2}
(Color online)
Hydrogen microstructure in a model of a-Si:H with 22\% H atoms showing a
cluster of H atoms (in yellow/white) consists of 6-10 atoms. A representative
SiH$_4$ configuration is also shown with the Si atom in green/white.  The other
configurations present in the environment include isolated H atoms, SiH
and SiH$_2$.
}
\end{figure}

We now consider the model with 22\% hydrogen in the network. This value is somewhat high for the model
to be considered as device-quality, nonetheless our intention is to study how the addition of more hydrogen
changes the microstructure in the network. Experimentally it has been observed that with increasing concentration
hydrogen atoms tend to form clusters until the concentration is high enough when small clusters are no longer
visible \cite{Gleason}. In fig.\ref{fig2} we have shown the microstructure of a region of space having linear
dimension of 6 {\AA}. The clustering of hydrogen atoms surrounding an SiH$_4$ configuration is explicitly shown 
in the figure. 
The local environment of this Si atom consists of 6-8 H atoms, and both monohydride and
dihydride configurations are realized in this model. In addition to these,  some isolated H
atoms are also found to exist in the network.  The entire region consists of about total 20 H atoms showing
both clustered and dispersed H atoms. Further analysis of different regions in the same model suggests 
that monohydrides can reside in both sparse and small clustered environment. The H-H neighbor distance 
in this model is found to be in the range of 1.6 - 2.4 {\AA} and the distribution is shown in fig.\ref{fig5}.
A closer inspection of the local structure (with a cut-off 3.5 {\AA}) reveals that higher values 
at 2.1 {\AA} and 2.35 {\AA} mostly originate from SiH$_2$ and SiH$_4$ configurations. This can be easily understood 
considering the bonding geometry of these configurations and taking into account of Si-H bond length and H-Si-H 
bond angle distributions. The microstructure of the entire model is
presented in fig.\ref{fig4} showing hydrogen configurations that are realized in our model. This clearly
demonstrates the inhomogeneous nature of the microstructure of the model.

\section{NMR linewidths of model a-Si:H}
In the preceding section we have discussed how the microstructure changes with increasing hydrogen content of the
sample. Since the local environment of H atoms is reflected in the shape and the
width of the NMR spectrum, one can obtain valuable information on microstructure from a knowledge of
these quantities. However, direct calculation of NMR spectra is very difficult from first-principles,
and one usually resorts to the approximation that the spectrum can be expressed as a convolution of
Lorentzian and Gaussian distributions. NMR experiments on variously prepared samples at different concentrations
reveal that at very dilute concentration randomly dispersed spins (via dipolar interaction) give rise to
a narrow line width, whereas the presence of small clusters can be best attributed to the broader linewidth
of the spectrum. These narrow and broad widths are generally assumed to be associated with the Lorentzian and Gaussian
broadening of the spectrum respectively. The concentration of hydrogen in our model is sufficiently
high to assume that the spectrum can be approximated by a Gaussian distribution to estimate the broad linewidth.  
This can be confirmed via calculation of $\Gamma$, the ratio of the fourth moment $(\mu_4)$ to the square of
the second moment ($\mu_2$) for the model networks and compare to the same with a Gaussian distribution. 
A value close to 3 (for ideal Gaussian) suggests that the spectrum can be approximated via Gaussian
lineshape for which the FWHM is given by $\sigma = 2.36 \,\sqrt{\mu_2}$ \cite{nmr-book}. For an amorphous
system with a random distribution of spins, the second and the fourth moments are given by \cite{nmr-book}:

\begin{eqnarray}
\label{eq2}
\mu_2 &=& -\frac{Tr[H,I_x^2]}{Tr[I_x^2]} \nonumber \\
&=& \frac{3}{4N} \gamma^4\hbar^2\, I(I+1)\, \sum_{j \ne k}^N \frac{(1-3\cos^2\theta_{jk})^2}{r_{jk}^6}
\end{eqnarray}
and

\begin{equation}
\label{eq3}
\mu_4 = \frac{Tr[H,[H,I_x]]^2}{Tr[I_x^2]}
\end{equation}

where N is the number of spins, I = $\frac{1}{2}$ and $\theta_{jk}$ is the angle between the vector ${r_{jk}} $ with
the applied magnetic field direction. A somewhat less accurate but simplified expression for $\sigma$ is given
by \cite{Rutland},
\be
\label{eq4}
\sigma_B({\rm kHz}) = 190 \left(\frac{1}{N} \sum_{i\ne j }^N \frac{1}{r_{ij}^6} \right)^{\frac{1}{2}}
\ee
where $r_{ij}$ is the distance between the spins in the unit of {\AA}.  Similarly, an estimate of the narrow linewidth
of the spectrum coming from the dispersed H atoms can be obtained from\cite{nmr-book}, 
\be
\label{eq5}
\sigma_N = \frac{4\, \pi^2}{3 \sqrt{3}}\, \gamma^2\hbar \, n
\ee
where n is the number density of the spins without counting those in the clustered environment.

\begin{figure}[t] 
\includegraphics[width=2.8in, height=2.3in, angle =0 ]{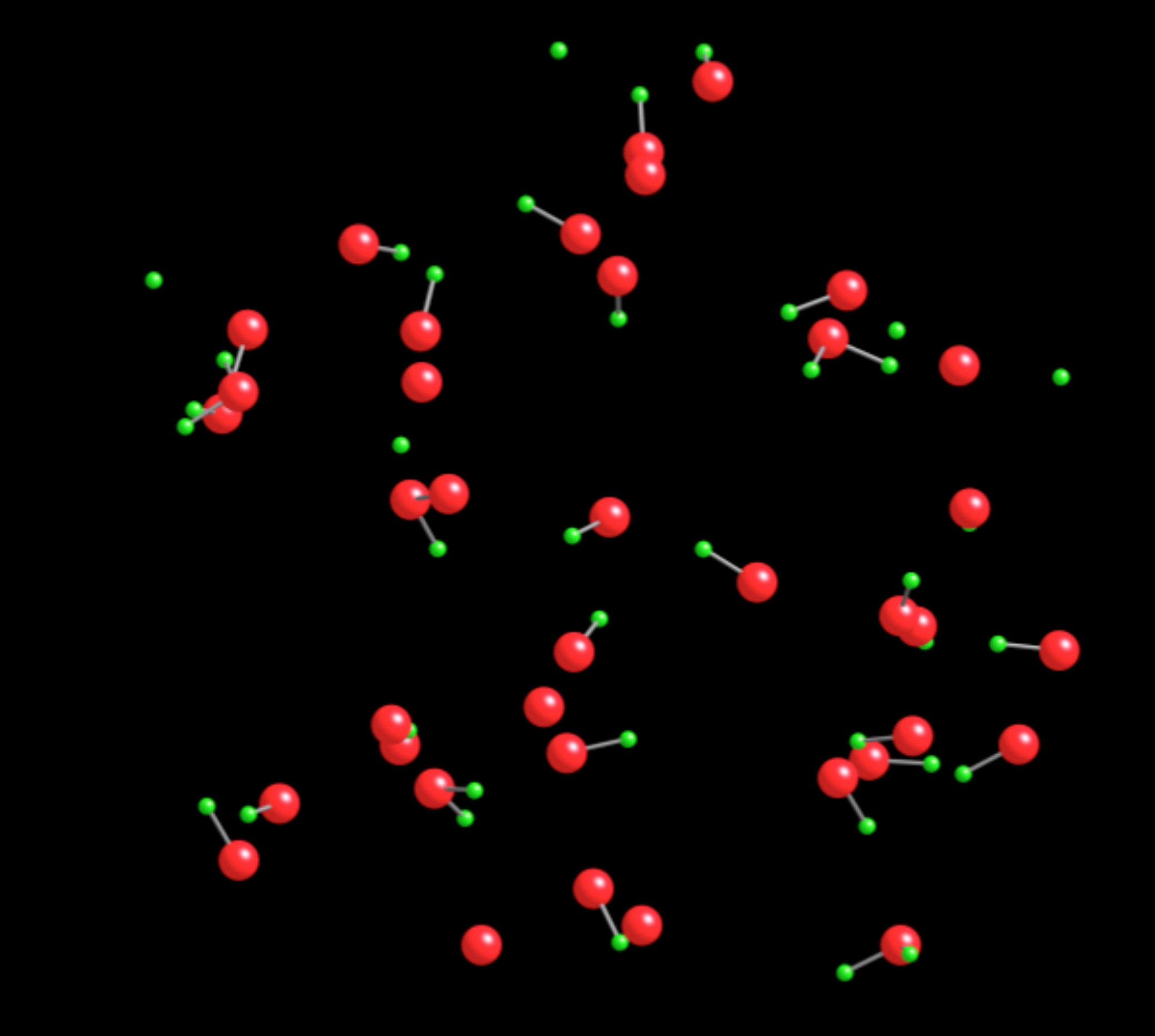}
\caption{
\label{fig3}
(Color online)
The overall hydrogen microstructure of the model with 7\% H obtained from the
ECMR method. The Si matrix is removed (except Si-H  bonds) in order to display
the nature of microstructure at low concentration. The Si and H atoms are
shown in red (dark grey) and green (light grey) respectively.
}
\end{figure}

\begin{figure}[t]
\includegraphics[width=2.8in, height=2.3in, angle =0 ]{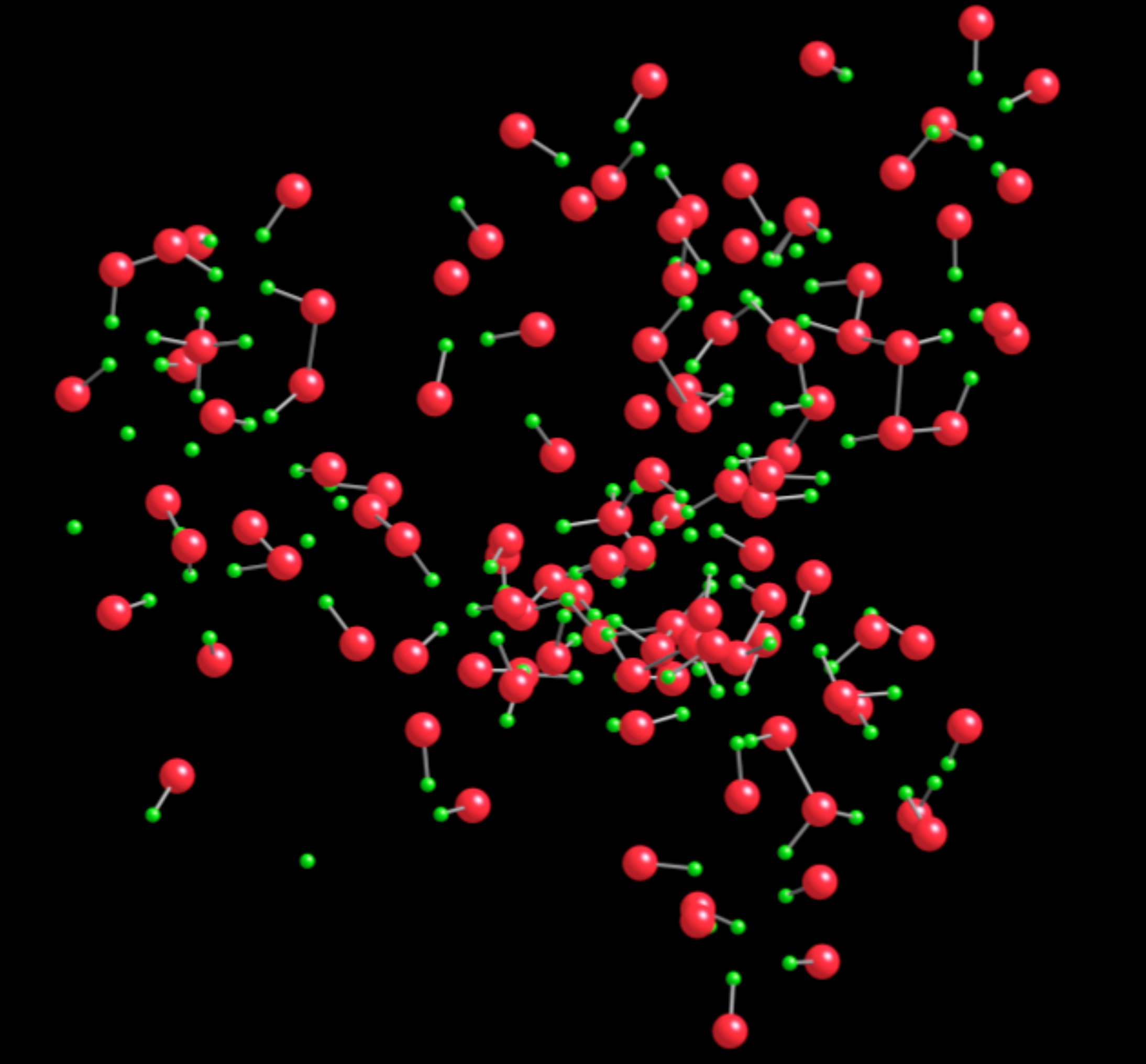}
\caption{
\label{fig4}
(Color online) The overall hydrogen microstructure of the model at high concentration
(22\% of H). Once again Si (in red/dark grey) and H (in green/light grey) atoms are displayed for clarity.
Several H clusters are present along with few isolated H atoms, SiH and SiH$_2$
configurations. Note that the microstructure consists of a combination of sparse and
clustered environment that are observed experimentally.
}
\end{figure}

\begin{figure}[t]
\includegraphics[width=3.3in, height=3.3 in, angle =0 ]{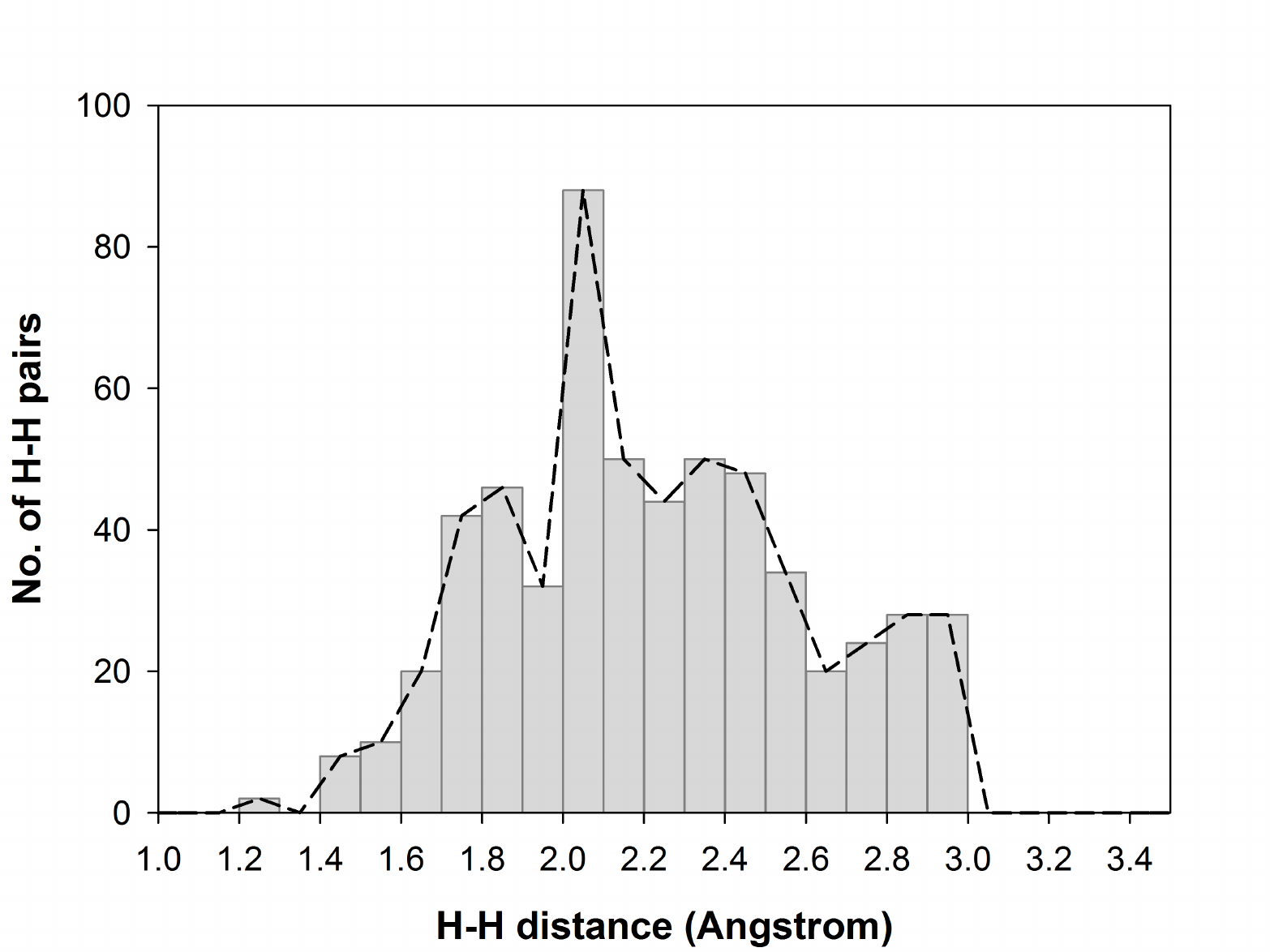}
\caption{
\label{fig5}
The distribution of H-H distances in a model of a-Si:H with 22\% H atoms. A cut-off
of 3.5 {\AA} is used to obtain the distribution. The peak at 1.8 {\AA} is found to originate
from the small clustered environment, whereas the peaks at 2.1 and 2.35 {\AA} can be shown to appear
from geometrical consideration of SiH$_2$ and SiH$_4$ configurations. 
}
\end{figure}

In table I, we have listed the ratio of the moments, $\Gamma$ and both the narrow and broad linewidths 
(FWHM) for the models for two different concentration. The average value of the moment is obtained from 
Eq.\ref{eq4} and is indicated in the table. The value of $\Gamma$ is found to be close but less 
than 3 in both the cases indicating the inhomogeneous nature of the distribution. 
This is expected in view of the inhomogeneous microstructure that we have seen in figs.\ref{fig1} and \ref{fig3}. These values are consistent
with those reported in the literature \cite{Street} \cite{Rutland}. The broad linewidth associated with 22\% H concentration
is of the order of 50kHz, which is somewhat high but is expected here because of the presence of a large cluster (cf. fig.~\ref{fig2})
of linear dimension 6.0 {\AA}. Our results support the recent observation of a significantly large linewidth in
hot-filament-assisted CVD deposited films by Wu et al.~\cite{Rutland}. On the basis of their NMR data,
these authors suspected the presence of large H clusters in the sample with few H nearest neighbors at a
distance 1.6 - 1.8 {\AA}.  Together with the microstructure, one can infer that large H complexes (with 
20 H atoms) are realizable, and when present they can produce a very broad linewidth in the spectrum.

\begin{table}[b]
  \caption{
NMR linewidths for model a-Si:H 
}
  \begin{tabular}[htbp]{@{}lllll@{}}
    \hline
    Conc ($\%$) & Directions & $\sigma_B$ (kHz) & $\sigma_N$ (kHz) & $\Gamma$ \\
    \hline
    7  & 001 & 16.34 (17.33) & 2.8 & 2.8  \\
         & 010 & 19.68 (17.33) & 2.8 & 2.6  \\
         & 100 & 17.80 (17.33) & 2.8 & 2.8  \\
    \hline
    22  & 001 & 52.38 (50.36)  & 6.4 & 2.9  \\
         & 010 & 52.15 (50.36) & 6.4 & 2.9  \\
         & 100 & 49.92 (50.36) & 6.4 & 2.9  \\
    \hline
  \end{tabular}
  \label{onecolumntable}
\end{table}

\section{Conclusion} 
The hydrogen microstructure in two realistic models of a-Si:H is studied at low and high
concentration of H atoms. The distribution of H atoms in both the models are found to be
inhomogeneous that consists of few isolated H atoms, monohydrides and dihydrides that 
reside in both sparse and clustered environment. In addition, few SiH$_4$ configurations 
are found to 
realize in our first-principles model networks. At low concentration, the microstructure essentially
consists of few small clusters of 4-5 atoms and a sparse distribution of H configurations.
For the network with high concentration of hydrogens, a cluster containing as many as 
20 H atoms has been observed in a region of linear dimension 6 - 8 {\AA}. The theoretical linewidths obtained
from the models using Gaussian and truncated Lorentzian broadening match with experimental
data.  In particular, our study confirms the presence of large clusters of hydrogen 
with 1.8 - 2.4 {\AA} as H-H neighbor distances. The existing multiple quantum NMR data 
confirm our results.

\begin{acknowledgement}
PB acknowledges the support of the University of Southern Mississippi under 
Grant No. DE00945 and Aubrey Keith Lucas and Ella Ginn Lucas Endowment for awarding 
a fellowship under faculty excellence in research program.

\end{acknowledgement}

%

\end{document}